\documentclass[superscriptaddress,aps,amsmath,amssymb,floatfix,reprint,raggedbottom,longbibliography]{revtex4-1}
\usepackage{graphicx}
\usepackage[colorlinks=true,citecolor=blue,linkcolor=blue,urlcolor=blue]{hyperref}
\usepackage{dcolumn}
\usepackage{xcolor}
\usepackage{fancyhdr}
\usepackage{lipsum}
\usepackage{soul}
\usepackage{braket}
\usepackage[utf8]{inputenc}
\DeclareUnicodeCharacter{2009}{\,} 

\makeatletter 
\renewcommand{\fnum@figure}{\textbf{FIG.~\thefigure}}
\makeatother

\makeatletter
\def\bbordermatrix#1{\begingroup \m@th
  \@tempdima 4.75\p@
  \setbox\z@\vbox{%
    \def\cr{\crcr\noalign{\kern2\p@\global\let\cr\endline}}%
    \ialign{$##$\hfil\kern2\p@\kern\@tempdima&\thinspace\hfil$##$\hfil
      &&\quad\hfil$##$\hfil\crcr
      \omit\strut\hfil\crcr\noalign{\kern-\baselineskip}%
      #1\crcr\omit\strut\cr}}%
  \setbox\tw@\vbox{\unvcopy\z@\global\setbox\@ne\lastbox}%
  \setbox\tw@\hbox{\unhbox\@ne\unskip\global\setbox\@ne\lastbox}%
  \setbox\tw@\hbox{$\kern\wd\@ne\kern-\@tempdima\left[\kern-\wd\@ne
    \global\setbox\@ne\vbox{\box\@ne\kern2\p@}%
    \vcenter{\kern-\ht\@ne\unvbox\z@\kern-\baselineskip}\,\right]$}%
  \null\;\vbox{\kern\ht\@ne\box\tw@}\endgroup}
\makeatother

\begin{document}
\title{Double Free-Layer Magnetic Tunnel Junctions for Probabilistic Bits }
\author{Kerem Y. Camsari}
\affiliation{Department of Electrical and Computer Engineering, University of California, Santa Barbara, Santa Barbara, CA, 93106, USA}
\author{Mustafa Mert Torunbalci}
\affiliation{Birck Nanotechnology Center, Purdue University, West Lafayette, IN, 47907, USA} 
\author{William A. Borders}
\affiliation{Laboratory for Nanoelectronics and Spintronics, Research Institute of Electrical Communication, Tohoku University, Japan}
\author{Hideo Ohno}
\affiliation{Laboratory for Nanoelectronics and Spintronics, Research Institute of Electrical Communication, Tohoku University, Japan}
\affiliation{Center for Spintronics Research Network, Tohoku University, Japan}
\affiliation{Center for Science and Innovation in Spintronics, Tohoku University, Japan}
\affiliation{WPI-Advanced Institute for Materials Research, Tohoku University, Japan}
\author{Shunsuke Fukami}
\affiliation{Laboratory for Nanoelectronics and Spintronics, Research Institute of Electrical Communication, Tohoku University, Japan}
\affiliation{Center for Spintronics Research Network, Tohoku University, Japan}
\affiliation{Center for Science and Innovation in Spintronics, Tohoku University, Japan}
\affiliation{WPI-Advanced Institute for Materials Research, Tohoku University, Japan}
\date{\today}
\begin{abstract}
Naturally random devices that exploit ambient thermal noise have recently attracted attention as hardware primitives for accelerating probabilistic computing applications.  
One such approach is to use a  low barrier nanomagnet as the free layer of a magnetic tunnel junction (MTJ) whose magnetic fluctuations are converted to resistance fluctuations in the presence of a stable fixed layer. Here, we propose and theoretically analyze a magnetic tunnel junction with no fixed layers but two free layers that are circularly shaped disk magnets.  We use an experimentally benchmarked model that accounts for finite temperature magnetization dynamics, bias-dependent charge and spin-polarized currents  as well as the dipolar coupling between the free layers. We obtain analytical results  for statistical averages of fluctuations that are in good agreement with the numerical model. We find that the free layers with low diameters fluctuate to randomize the resistance of the MTJ in an approximately  bias-independent manner. 
We show how such MTJs can be used to build a binary stochastic neuron (or a p-bit) in hardware.  Unlike earlier stochastic MTJs that need to operate at a specific bias point to produce random fluctuations, the proposed design can be random for a wide range of bias values, independent of spin-transfer-torque pinning. Moreover, in the absence of a carefully optimized stabled fixed layer, the symmetric double-free layer stack can be  manufactured using present day Magnetoresistive Random Access Memory (MRAM) technology by minimal changes to the fabrication process. Such devices can be used as hardware accelerators in energy-efficient computing schemes that require a large throughput of tunably random bits. 
\end{abstract}
\maketitle
%\thispagestyle{fancy}
%\listoffigures
%\listoftables
%\tableofcontents

\section{Introduction}

Intrinsic randomness in nanodevices can be harnessed to do useful computational tasks, especially when the natural physics of a device map to a useful functionality, a principle sometimes expressed as ``let physics do the computing'' \cite{parihar2017computing}. One such example is the physics of low-barrier magnets that can produce random fluctuations
in magnetization which can be turned into fluctuations in resistances in Magnetic Tunnel Junctions (MTJs). Stochastic MTJs (sMTJ) and the stochastic behavior of MTJs have attracted a lot of  recent attention both theoretically and experimentally \cite{locatelli2014spin,fukushima2014spin,choi2014magnetic,kim2015spin,lee2017design,majetich_2018,vodenicarevic_low-energy_2017,vodenicarevic_circuit-level_2018,lv2017single,liyanagedera2017stochastic,debashis2018tunable,mizrahi2018neural,debashis2018tunable,borders2019integer,ostwal2019spin,abeed2019low,daniels2020energy,parks2020mr,ganguly2020building,safranski2020demonstration,PhysRevApplied.11.034015}. It has been observed that even with relatively modest tunneling magnetoresistance of present day MTJs where the parallel to anti-parallel resistance ratios are small, fluctuations in resistances can be converted to electrical fluctuations that can be sensed by inverters or amplifiers \cite{camsari2017implementing,borders2019integer}.  Such devices can be useful as compact, energy-efficient tunable true random number generators that can be interconnected to accelerate a wide range of computational tasks such as sampling and optimization \cite{camsari_pbits_2019}.  

Both theory and available experimental data suggest that when sMTJs are designed out of perpendicular easy axis MTJs (p-MTJ) where both the fixed and the free layer have perpendicular anisotropy, the fluctuations tend to be slow since even at the zero-barrier limit $E / k_B T \rightarrow 0$ ($E$ is the energy barrier of the magnet and $k_B$ is Boltzmann constant and $T$ is temperature) fluctuations ($\tau$) are of the order of $\tau^{-1} \approx \alpha \gamma H_{th}$ ($\gamma$ is the gyromagnetic ratio of the electron, $\alpha$ is the damping coefficient and $H_{th}$ is an effective thermal noise field given by $H_{th}=k_B T/M_s \mathrm{Vol.}$, $M_s \mathrm{Vol.}$ being the total magnetic moment)  \cite{coffey2012thermal}. Even for a small magnet, for example with half a million spins, $\tau^{-1}$ is limited to frequencies around 1.5 MHz $-$ 15 MHz  for $\alpha=0.01 - 0.1 $.   An alternative  is to use circular disk magnets with no intrinsic anisotropy since  by virtue of their large demagnetizing field, these magnets tend to produce much faster fluctuations \cite{camsari2017stochastic,camsari2017implementing} with a precessional mechanism that has been theoretically analyzed \cite{kaiser2019subnanosecond,hassan2019low,kanai2021} and recently observed in experiment \cite{safranski2020demonstration,hayakawa2021}. All focus on sMTJs, whether with in-plane or perpendicular easy-axis magnets,  however, has been on magnetic stacks where there is a stable fixed layer and an unstable free layer that fluctuates in the presence of thermal noise. The current standard in spin-transfer-torque magnetoresistive random access memory (STT-MRAM) technology is to use p-MTJs \cite{bhatti2017spintronics} and switching to such in-plane easy-axis magnets with at least one stable fixed layer to get fast fluctuations is challenging from an industry standpoint, especially for miniaturized MTJs down to a few tens of nanometers where techniques to fabricate fixed layers have not been established.

\begin{figure}[!t] 
\centerline{\includegraphics[width=1\linewidth]{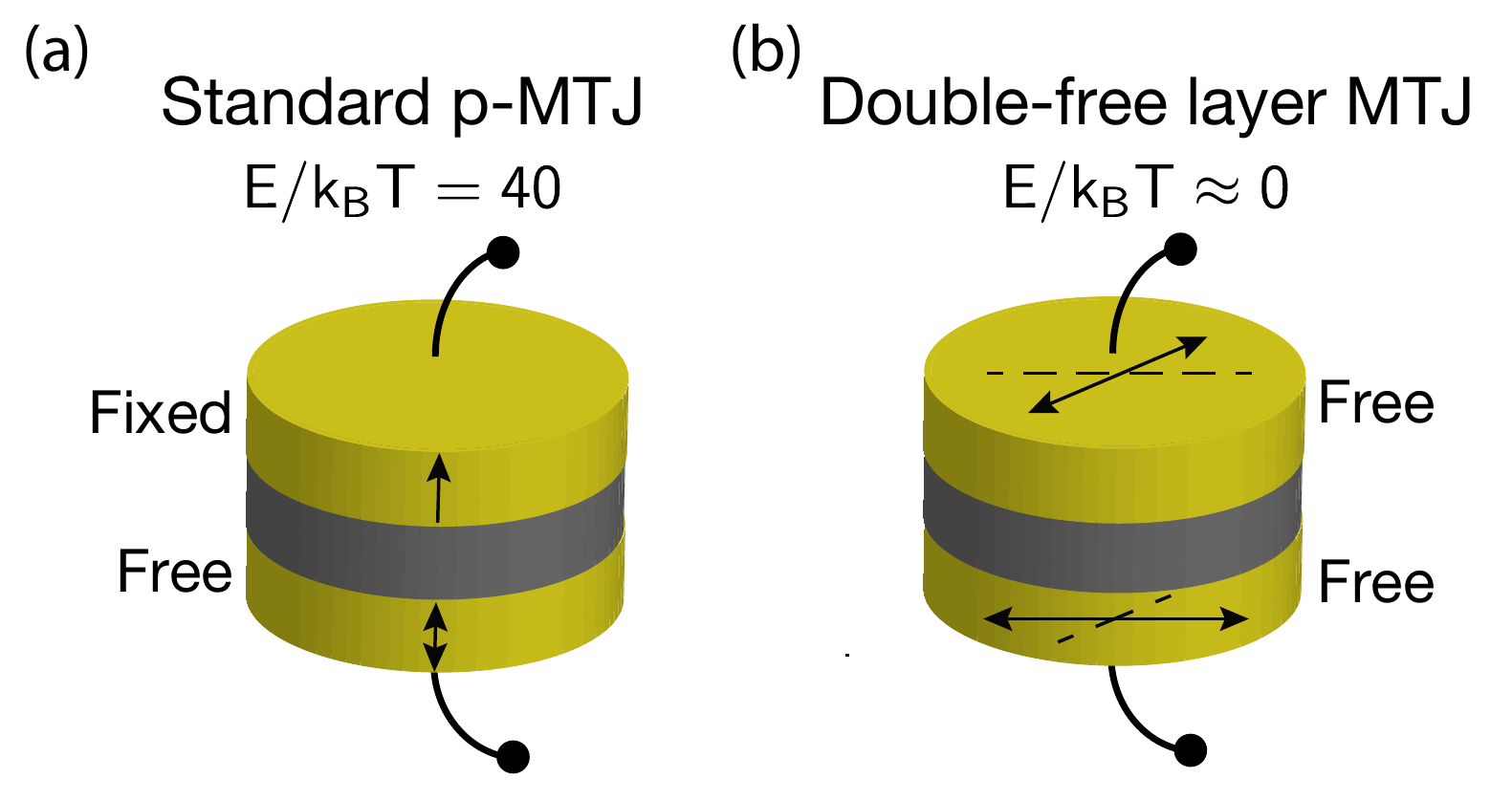}}
\caption{\textbf{Proposed device:} (a) Standard MTJ of the MRAM technology using perpendicular easy axis MTJs (p-MTJ) with fixed and free layers. (b) We consider an MTJ with two free layers and no fixed layer. The magnetization of the free layers fluctuate in the plane in the presence of thermal noise that are turned into resistance fluctuations through tunneling magnetoresistance (TMR). One way to build the proposed device is to start from the standard PMA-MTJs and make both free and fixed layers thicker so that their magnetizations fall into the plane. We note that the p-MTJ cartoon stack shown here is for illustrative purposes and industrial p-MTJ stacks have many more additional layers that account for canceling dipolar fields, ensuring fixed layer stability and other effects.}
\label{fi:fig1}
\end{figure}

In this paper, we propose and evaluate the possibility of using a  \textit{double free layer} MTJ where both layers are designed as circular in-plane easy-axis magnets. Such a configuration can be easily achieved by starting from a typical STT-MRAM material stack structure comprised of CoFeB/MgO/CoFeB MTJs  by making both the free and fixed layer magnets thicker such that their easy-axis orients in the plane of the magnet (FIG.~\ref{fi:fig1}). In the rest of this paper, we analyze the behavior of this double-free layer magnetic tunnel junction device. One advantage of this device comes from its simplicity: It is a completely symmetric device with two free layers that are in an in-plane configuration in equilibrium  and this does not require a highly optimized magnetic stack design as it is based on the same CoFeB/MgO/CoFeB structure of standard p-MTJs. Another key feature of this device is its bias-independence over a wide range of voltages, which can be useful for designing devices to be used in probabilistic computing applications where a large throughput of tunable random bitstreams are needed. We note that double free layer structures similar to those shown in FIG.~\ref{fi:fig1} have been discussed in the context of spin-torque nano oscillators (see for example Ref.~\cite{kudo2006synchronized,rowlands2012magnetization,taniguchi2019synchronized,matsumoto2019chaos,zhou2019inducing}), however our focus in this paper is on fully circular magnets  with no intrinsic anisotropy that are in the superparamagnetic regime.  

%  the unstable free layer cannot be magnetized in the plane to exploit fast fluctuations because with a perpendicularly magnetized fixed layer, in-plane fluctuations would not  be reflected in the MTJ resistance, due to a constant 90$^\circ$ angle between the layers. On the other hand, the industrial trend in STT-MRAM is to use perpendicular MTJs both for fixed and free layers and substantial changes in stack design may prove difficult to realize scalable sMTJs. 

The rest of the paper is organized as follows: In Section \ref{sec:magnetostatics} we develop a model to describe the dipolar interaction between the layers, starting from Maxwell's equations in the magnetostatics regime. In Section \ref{sec:llg}, we describe the finite temperature coupled macrospin model that describes the magnetization dynamics. In Section \ref{sec:zb}, we analyze the zero-bias behavior of the double free layer MTJ with analytical benchmarks that are obtained from equilibrium statistical mechanics. In Section \ref{sec:fb}, we describe the fully voltage dependent model that considers bias-dependent spin-polarized currents that influence the free layers. Finally in Section~\ref{sec:pbit}, we show how the proposed device can be combined with modern transistors in a 1T/1MTJ circuit topology to deliver tunable randomness with fast fluctuations. 
\section{Magnetostatics} 

\label{sec:magnetostatics}
In the absence of any external magnetic field and intrinsic anisotropies, the energy of the 2-magnet system is fully specified by  magnetostatics and is given by \cite{mayergoyz2009nonlinear}:
%\vspace{-25pt}
\begin{equation}
E\hspace{-2pt} = -2 \pi M_s^2 \mathrm{Vol.} \hspace{-4pt} \left(\sum_{i=1}^{2} \hat{m}_i^T {N}_{ii} \hat{m}_i \hspace{-0pt}+\hspace{-0pt} \sum_{\substack{i,j \\ i \neq j}}^{2} \hat{m}_i^T {D}_{ij} \hat{m}_j\right) 
\label{eq:energy}
\end{equation}

where $M_s$ is the magnetic moment per volume, $\mathrm{Vol.}$ is volume, $\hat{m}_i$ are the 3-component magnetization vectors,
 $N_{ii}$ and $D_{ij}$ are the demagnetization and the dipolar tensors, respectively (we adopt cgs units for magnetic models throughout). We have implicitly assumed the volume and the $M_s$ to be the same for both magnets,
 which is true for all cases considered in this paper.  We adopt a macrospin approach where the chosen volume corresponds to the volume of the magnet and $\hat{m}_i$ are described as 3-component vectors. We numerically solve for $\textbf{D}$ and $\textbf{N}$ 
 that are in general position dependent, but we average them within the volume of the target magnet to reduce these to single numbers for a given geometry.

\begin{figure}[!t] 
\centerline{\includegraphics[width=0.99\linewidth]{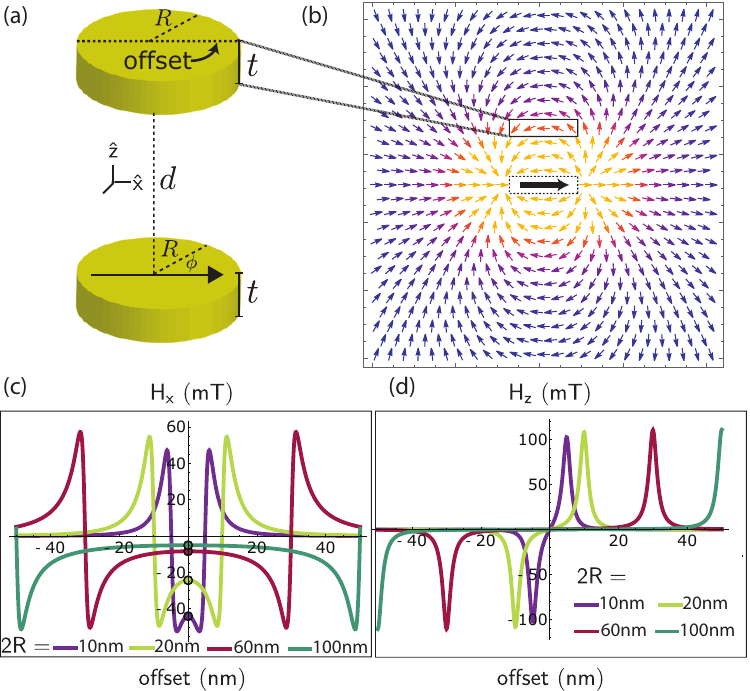}}
\caption{\textbf{Dipolar model:} (a) The geometry and parameters in the calculation of dipolar fields between the two free layers. (b) Illustrative vector plot of the dipolar field due to the bottom layer (dashed box) at y=0 plane. (c) The dipolar field $H_X$ due to a source magnet (bottom, +$x$ polarized) with varying diameters ($2 R$=$10, 20, 60, 100$ nm) with a film thickness of $t$=1 nm at a distance $d$=1 nm (typical MgO thickness in MTJs)  measured from the top of the source magnet with $M_s$ = 800 emu/cc $\approx 1$ T. At zero offset (circled points) the fields can be analytically calculated from Eq.~\ref{eq:simple}. (d) $z-$directed fields along the offset direction. In our model, these fields average out to zero when summed over the target magnetic volume.}
\label{fi:dip}
\end{figure}

Starting from the magnetostatics conditions where $\vec{\nabla}\times \vec{H}=0$, we can define a magnetic potential $\Phi$ such that 
$-\nabla\Phi = \vec{H} $ and since $\vec{\nabla}\cdot \vec{B}=0$ and $\vec{B}= (\vec{H}+4\pi \vec{M})$ always hold, we obtain $\nabla^2 \Phi = 4\pi  \ \vec{\nabla}\cdot \vec{M}$, 
which is mathematically equivalent to the Poisson equation of electrostatics. To solve this equation we first introduce a Green's function, $G(\vec{r},\vec{r'})$,
defined as  the potential at $\vec{r}$ due to a ``unit'' charge at source $\vec{r'}$.  $G$ can be readily identified from the definition of the Dirac-delta function \cite{arfken1999mathematical}, $\nabla^2 (-1/|\vec{r}-\vec{r'}|)\equiv 4\pi \delta(\vec{r}-\vec{r'})$.
Once this Green function is known, using the linearity of the potential \cite{purcell1965electricity}, we can write the general solution for the magnetic potential as
\begin{equation}
\Phi (\vec{r})= \displaystyle \int d\vec{r'} G(r,\vec{r'}) \rho_M(r') 
\end{equation}
 where we have defined a magnetic source density, $\rho_M(r') =  4\pi \vec{\nabla}\cdot \vec{M}$ which is only non-zero at the boundaries of the magnetic volume,  assuming a uniformly
 magnetized body. The solution to the magnetic potential at position $\vec{r}$ then  becomes:
 \begin{equation}
 \Phi (\vec{r}) = \int_{\mathrm{Vol.}} \frac{-\vec{\nabla}\cdot \vec{M}}{ |\vec{r}-\vec{r'}|}d\Omega
 \end{equation}
 where $\mathrm{Vol.}$ is the volume of the source magnet. Now let us consider the specific case shown in FIG.~\ref{fi:dip}a where we have a cylindrical in-plane magnet, $\hat M = M_s \hat x$.
The magnetic source density can be expressed as: $- \vec{\nabla}\cdot \vec{M} = M_s \delta(R-r')\cos\phi$ since the magnetization
 abruptly becomes zero right outside the magnetic boundary. This allows us to write the magnetic potential as:

  \begin{equation}
  \Phi (\vec{r})\hspace{-2pt} = \hspace{-2pt}\int\displaylimits_{\mbox{-}\delta}^{\delta}\hspace{-2pt}\int\displaylimits_{0}^{2\pi}\hspace{-2pt}\frac{d\phi \ dz' \cos(\phi) M_s R}{\sqrt{(x\hspace{-2pt}-\hspace{-3pt}R \cos \phi)^2\hspace{-2pt}+\hspace{-2pt}(x\hspace{-2pt}-\hspace{-2pt}R \sin\phi)^2\hspace{-2pt}+\hspace{-2pt}(z\hspace{-2pt}-\hspace{-2pt}z')^2}} 
  \label{eq:int}
 \end{equation}
\normalsize

where we introduced $\delta=t/2$, $t$ being thickness of the magnetic layer (source).  We have not found a closed form solution of this integral, though it can be partly integrated just along  $z'$  after taking derivatives of $\Phi$ to obtain field expressions. These are not necessarily
informative so we do not repeat them here but we used these partial integrals to ease the numerical integration of Eq.~\ref{eq:int} (for another treatment, see for example, \cite{taniguchi2018analytical}). FIG.~\ref{fi:dip}b shows a typical position dependent vector plot of this numerical integration. In FIG.~\ref{fi:dip}c, we show typical results where the field strength increases for magnets with lower diameters, also observed in experiments with perpendicular MTJs \cite{gajek2012spin} (Also see inset FIG.~\ref{fi:zero_bias}a for averaged out $D_{xx}=D_0$ values). This correspondence between p-MTJs and the double free layer system considered here can be seen from the properties of the dipolar tensor, i.e., $\mathrm{tr}. [\mathbf{D}]=0$ \cite{newell1993generalization,wysin2015magnetic}. 

As mentioned earlier, our approach of calculating dipolar tensor components is by averaging field over the target magnetic body, for example to compute $D_{xx}$, we first compute $H_X$ over the target magnet volume
when the source is magnetized along $+x$ and take an average of this field to obtain a single value for $D_{xx}$. Eq.~\ref{eq:int}  can be easily integrated on the cylindrical axis ($x=0,y=0$) after taking the derivative $H_Z = -\partial \Phi / \partial z$ and this results in:
\begin{equation}
H_Z = \hspace{-2pt} \pi M_s\hspace{-3pt} \left(\frac{z-\delta }{\sqrt{R^2+(z-\delta )^2}}-\frac{z+\delta}{\sqrt{R^2+(z+\delta)^2}}\right) 
\label{eq:simple}
\end{equation}
FIG.~\ref{fi:dip}c shows the application of this formula at zero offset which is defined as the dashed line in FIG.~\ref{fi:dip}a that passes along the $y=0$ line and this result matches the numerical integration. It might be tempting to use Eq.~\ref{eq:simple} to approximate the dipolar tensor coefficients analytically to obtain a single number but we found that
this tends to significantly differ from the value we obtain after averaging over the volume, especially at larger diameters. With this method of calculating dipolar coefficients, we find that $D_{xy}$, $D_{yx}$, $D_{zx}$, $D_{zy}$  all average to zero
which only leaves the diagonal components (see FIG.~\ref{fi:dip}d). Moreover, the cylindrical symmetry of the problem ensures $D_{xx}=D_{yy}=D_0$ and from the symmetry of the dipolar tensor $D_{zz} = -2 D_0$, leaving only one tensor coefficient to compute. Similarly,  the diameter ($2R$) to thickness
 ($t$) ratios  of all the magnets analyzed in this paper ensure $t/R \ll 1$, and the demagnetization tensor always has one component, $N_{zz}\approx -1$ which is what we use in the rest of the paper.

\section{Magnetization dynamics} 
\label{sec:llg}

\begin{figure}[!t] 
\centerline{\includegraphics[width=0.75\linewidth]{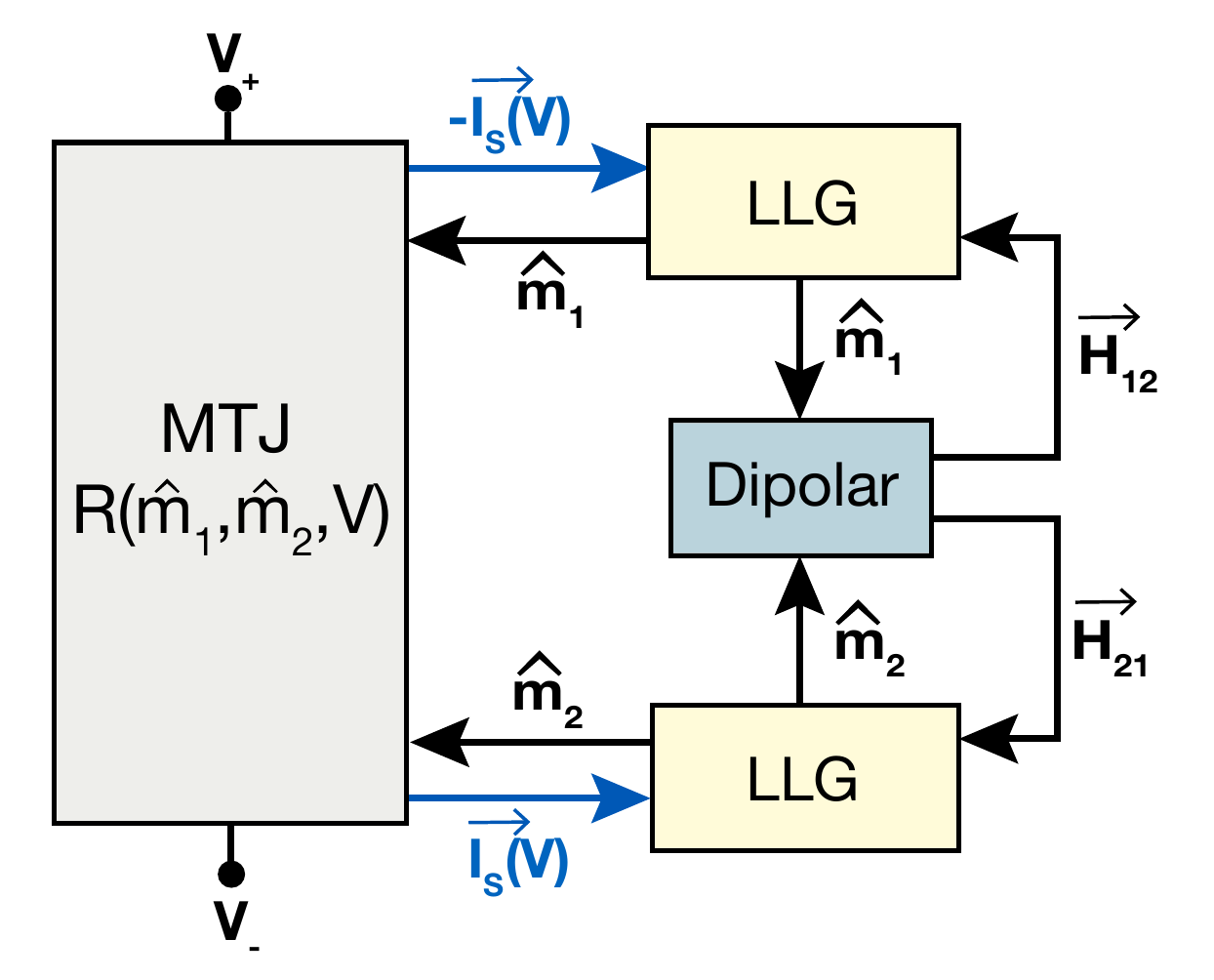}}
\caption{\textbf{Self-consistent model for the magnetization dynamics and transport:} The coupled Landau-Lifshitz-Gilbert (LLG) equations provide instantaneous magnetizations to the Magnetic Tunnel Junction (MTJ) model, which in turn produces a bias-dependent spin-polarized current, $\vec{I}_S (V)$ in the channel.  $+\vec{I}_S (V)$ is incident to one ferromagnetic interface and $-\vec{I}_S (V)$ is incident to the other ferromagnetic interface.  The dipolar model couples the two LLG solvers through fields that depend on instantaneous $\hat{m}_i$. }
\label{fi:model} 
\end{figure}

\begin{figure}[!t] 
\centerline{\includegraphics[width=0.99\linewidth]{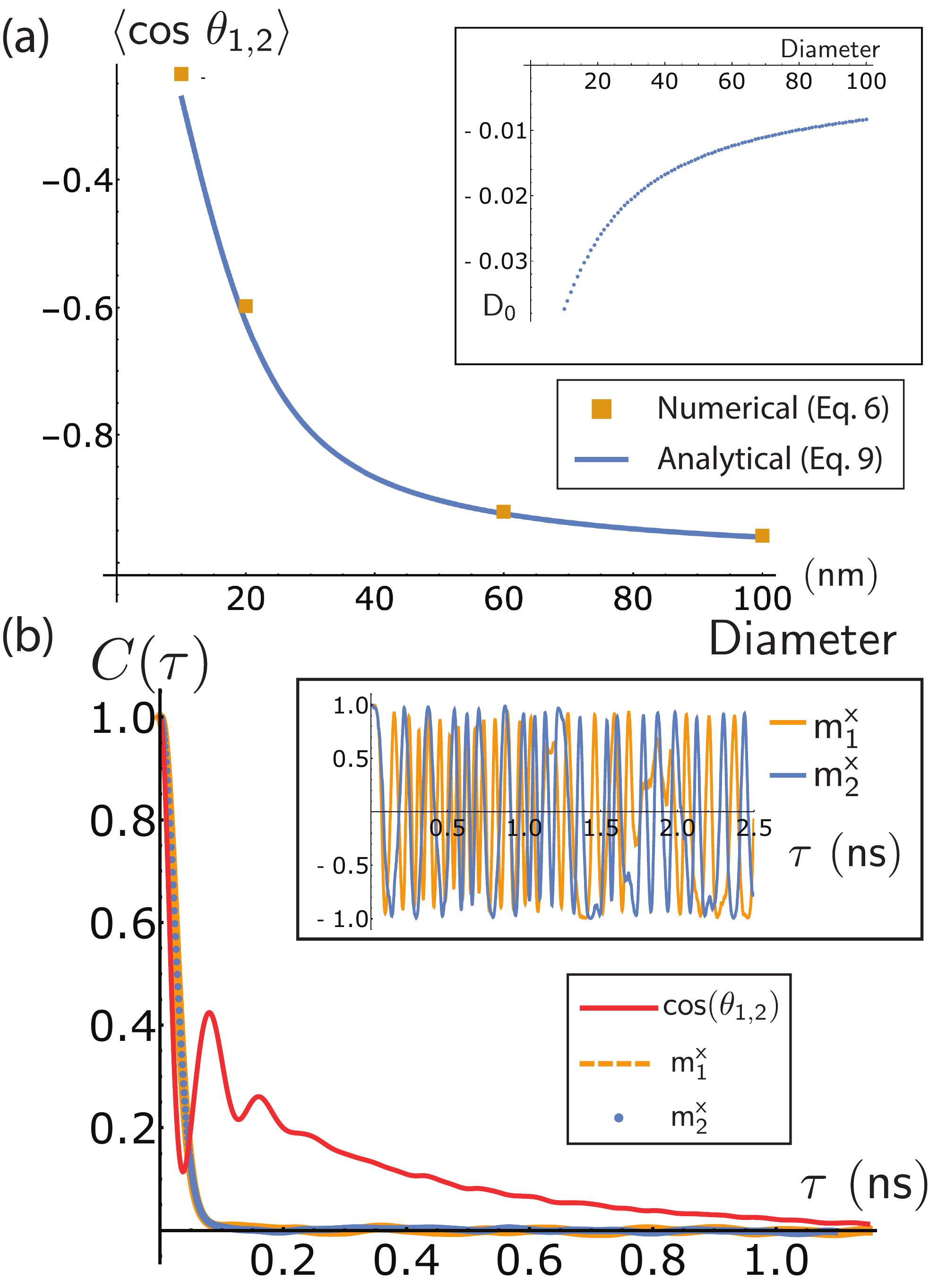}}
\caption{\textbf{Zero-bias behavior:} (a) The average $\cos\theta_{1,2}$ ($\theta_{1,2}$ is the angle between magnetizations vectors) at zero-bias. Eq.~\ref{eq:dipolarA} is compared with a finite temperature LLG simulation at different diameters (measured in nm). The average is taken over 5 $\mu$s for all examples with a time step of $\Delta t=1$ ps. (Inset) The average dipolar tensor component $D_{xx}=D_{yy}=D_0$ is shown as a function of diameter calculated from Eq.~\ref{eq:int} assuming the same geometric and magnetic parameters that were used in FIG.~\ref{fi:dip}, damping coefficient $\alpha=0.01$. (b) The normalized auto-correlation of $m^x$ for both layers and $\cos(\theta_{1,2})$   are shown for $2R=10$ nm as an example, where the total simulation period is 5 $\mu$s.  (Inset) Time dependence of $m^x$ components for a short period.}
\label{fi:zero_bias}
\end{figure}

Next, we describe the coupled magnetization dynamics model that considers two coupled Landau-Lifshitz-Gilbert (LLG) equations at finite temperature \cite{butler2012switching,sun2000spin,sun2004spin}:
\begin{eqnarray}
&& (1+\alpha^2)\frac{d\hat m_i}{dt} = -|\gamma|{\hat m_i \times \vec{H}_i} - \alpha |\gamma| (\hat m_i \times \hat m_i \times \vec{H}_i)\nonumber \\  && +  \frac{1}{q  N}(\hat m_i \times \vec{I}_{S_i}(V) \times \hat m_i)  + \left(\frac{\alpha}{q N} (\hat m_i \times \vec{I}_{S_i}(V))\right)
\label{eq:llg}
\end{eqnarray}
where $\alpha$ is the damping coefficient, $\gamma$ is the gyromagnetic ratio of the electron and $i$ is the magnet index, $i \in \{1,2\}$. The effective field for each magnet is calculated according to Eq.~\ref{eq:energy}, $H_i \equiv -\nabla_{\hat m} E/ (M_s \mathrm{Vol.})$ and  $N = (M_s \mathrm{Vol.}) / \mu_B$, and $\mu_B$ is the Bohr magneton.   The effective field $H_i$ includes uncorrelated Gaussian noise ($H^n_{x,y,z}$) at each direction $(x,y,z)$ with the following statistical properties: $\langle H^n(t) \rangle = 0$, $\langle H^n(t) H^n(t') \rangle =D \delta(t-t')$   where $D = (2 \alpha k_B T) /(\gamma M_s \mathrm{Vol.}$). In our model, this set of equations are solved self-consistently with a transport model for the MTJ, which provides the bias-dependent spin-polarized current, $\vec{I}_S(V)$ at a given bias, which in turn depends on the instantaneous magnetizations, $m_i$ (FIG.~\ref{fi:model}). 

An important consideration when modeling circular disk nanomagnets in the macrospin approximation is the formation of vortex states \cite{shinjo2000magnetic}. However, for the highly reduced diameters ($\le$100 nm) and thicknesses ($\le 1$ nm) considered in this paper, both detailed micromagnetic simulations \cite{moreira2017decreasing,lai2008size,ha2003micromagnetic} and available experiments \cite{cowburn1999single,debashis2016experimental,debashis2018design} indicate that the macrospin modeling approach in the parameter ranges considered should be reasonably informative.

We solve the stochastic LLG equation using the transient noise function of HSPICE. We have extensively benchmarked this model by comparing its time-dependent statistical behavior with respect to the Fokker-Planck Equation  \cite{torunbalci2018modular} and as well as by comparing it against our own implementation that solves the stochastic LLG using the Stratonovitch convention \cite{behin2010proposal}.

\section{Zero-bias behavior }
\label{sec:zb}

It is instructive to analyze the zero-bias behavior ($V=0 \rightarrow \vec{I}_S=0$) of the two free layers before considering their interaction with spin-polarized currents. 
Starting from Eq.~\ref{eq:energy} and making use of the symmetry results described at the end of Section \ref{sec:magnetostatics}, we can write:
\begin{eqnarray}
E= - 4\pi M_s^2 \mathrm{Vol.} [ -\frac{(m_1^z)^2}{2} - \frac{(m_2^z)^2}{2}  + \nonumber \\
(D_0) m_1^x m_2^x+(D_0) m_1^y m_2^y -(2 D_0) m_1^z m_2^z]
\label{eq:energy2}
\end{eqnarray}

Since both free layers are low-barrier magnets that fluctuate in the presence of thermal noise, it is instructive to calculate the cosine of the average angle between the free layers ($\cos\theta_{1,2}$)
at zero-bias as this determines the resistance of the MTJ. This average can be  written down from the Boltzmann distribution (switching to spherical coordinates):
\begin{equation}
\langle \cos\theta_{1,2} \rangle \hspace{-3pt}=\hspace{-3pt} \frac{1}{Z}\hspace{-3pt} \int \hspace{-3pt}\hat m_1\hspace{-2pt}\cdot\hspace{-2pt} \hat m_2\hspace{-1pt}\exp\hspace{-3pt}\left[\frac{-E(\theta,\phi,\eta,\chi)}{k_B T}\right] \hspace{-3pt}d\theta d\phi d\eta d\chi 
\label{eq:boltzmann}
\end{equation} 
where $(\theta,\phi)$, $(\eta,\chi)$ are spherical coordinate pairs for the two magnets and $Z$ is a normalization constant that ensures the total probability is 1. We cannot
find a  closed form expression for this integral but we can make progress by approximations.  Introducing $h_d \equiv 4 \pi M_s^2 \mathrm{Vol.}  / k_B T$ and $d_0 = D_0 h_d$, we note that
for typical parameters ($M_s$=800 emu/cc  and $2R=10-100$ nm), $h_d \gg 1$ indicating both magnets always roughly remain in the $(x,y)$ plane in equilibrium. Expressing this assumption mathematically, we expand
the integrand in Eq.~\ref{eq:boltzmann} at $(\theta,\eta)\rightarrow(\pi/2,\pi/2)$ and keeping the leading order terms, we obtain:
\begin{equation}
\langle \cos\theta_{1,2} \rangle\approx \frac{I_1(d_0)}{I_0(d_0)}
\label{eq:dipolarA}
\end{equation}
where $I_n$ is the modified Bessel function of the first kind. 

This simple expression determines the degree of coupling between the two layers in terms of the cosine of the angle
between their magnetization and is entirely dependent on geometric and material parameters.
FIG.~\ref{fi:zero_bias}a shows a comparison of Eq.~\ref{eq:dipolarA} with numerical simulation of Eq.~\ref{eq:llg} and we observe that the analytical expression reproduces the numerically observed
average with high accuracy, especially at higher diameters where our assumption of $h_d \gg 1 $ becomes more accurate. The slight deviation at low diameters (10, 20) nm is due to this assumption becoming inaccurate.
Eq.~\ref{eq:dipolarA} is an important result of this paper, since as we see in Section~\ref{sec:fb}, the average angle behaves in a bias-independent manner, therefore Eq.~\ref{eq:dipolarA} is approximately valid also in non-equilibrium condition.
By providing the average degree of coupling between the layers in terms of material and geometric parameters, Eq.~\ref{eq:dipolarA} could be useful in the design process of double-free layer magnetic tunnel junctions. 

We observe from FIG.~\ref{fi:zero_bias}a (inset) that even though the average dipolar interaction strength $D_0$ decreases at high diameters,  the coupling strength observed through the average angle between the magnetizations
increases (FIG~\ref{fi:zero_bias}a). The reason for this is that increasing the diameter increases volume and the dipolar coupling energy as $\propto R^2$  compared to the thermal energy $k_B T$, while the dipolar interaction scales as $\propto R^{-1}$.  In other words, the decrease
in the dipolar coupling at high diameters is not enough to compensate for the rapidly increasing dipolar energy.  Another interesting observation is made when we observe the autocorrelation:
\begin{equation}
 C(\tau)=\frac{1}{T_{p}} \int m^x(t) m^x(t+\tau) dt
 \end{equation}
 of two in-plane components $m^x$ (FIG.~\ref{fi:zero_bias}b), where $T_p$ is the simulation period.  The $m^x$ components of each layer lose memory very rapidly in around 100 ps but the cosine of the angle between the magnets takes about 1 ns 
to get completely uncorrelated. Since this is the parameter that determines the resistance of the MTJ, it is the more relevant time scale to consider to generate random signals.
Note that in our analysis, we did not include effects of exchange interaction between the layers  that might be present in real MTJs, as we assume MgO  produces a low degree of exchange coupling compared to the dipolar coupling \cite{yang2010effect} but this may
require further investigation. We have also ignored the effect of an existing interfacial anisotropy in our energy model. Existence of a strong interfacial anisotropy may be detrimental to the speed of fluctuations \cite{kaiser2019subnanosecond}, hence strategies to reduce it 
might be useful. 
 
\section{Voltage dependent behavior}
\label{sec:fb}

\begin{figure}[t] 
\centerline{\includegraphics[width=0.999\linewidth]{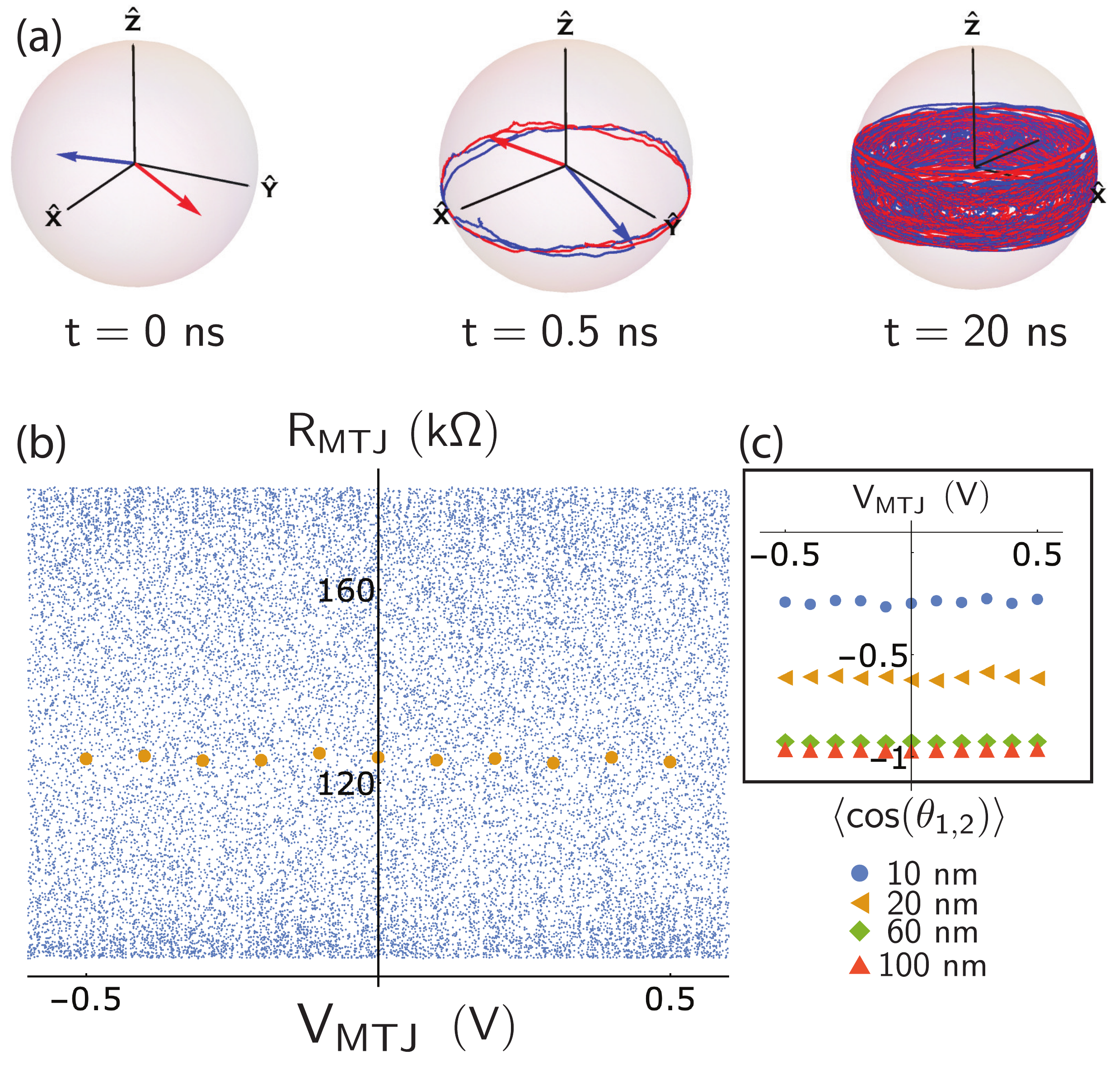}}
\caption{\textbf{R vs V characteristics of a double free layer MTJ:} (a) Representative fluctuations at different time instances for  $2R=20$ nm  free layers at t=0 ns, t=0.5 ns and t=20 ns.
 (b) $R$ vs $V$ characteristics for a $(2R)$=10 nm MTJ where we assumed a low-bias TMR of  $115\%$ ($P_0=0.65, V=0 $ in Eq.~\ref{eq:pol}), roll-of constant $V_0= 50$ V, ensuring a symmetric bias-dependence for the spin-polarized currents. We assume an $RA$-product of 9 $\Omega$-$\mu m^2$ \cite{lin200945nm} to obtain $G_0$ for all MTJs in this paper. The plot is obtained by sweeping the voltage from ($-$2V, 2V) in 2 $\mu$s with 1 ps time steps, dots correspond to 1 $\mu$s averages taken at that bias.  (c) Shows the average $\cos(\theta_{1,2})$ taken over 1 $\mu$s at different bias points for different diameters. All diameters roughly show bias-independent average angles or resistances.}
\label{fi:full_bias}
\end{figure}

In order to describe full bias dependence of the double-free layer MTJ, we first introduce our combined transport and magnetization dynamics model (FIG.~\ref{fi:model}). We describe the MTJ as a bias-dependent resistor that provides a bias-dependent
spin-polarized current to two separate finite temperature LLG solvers described by Eq.~\ref{eq:llg}. The LLG solvers provide magnetization vectors that change the resistance of the MTJ.  This self-consistency between the resistance that depends on magnetization and magnetization that depends on resistance is well-defined since magnetization dynamics are far slower than electronic time scales, as such, at each time point the resistance of the MTJ can be taken  as a lumped model that provides spin-polarized currents to the LLG solvers. Similarly the dipolar coupling acts ``instantaneously'',  providing updated fields that are fed back into the LLG solver at each time step.

We model the bias-dependent conductance ($G_{MTJ}\equiv 1/R_{MTJ}$)  based on two voltage-dependent interface polarizations \cite{datta2011voltage,datta2012modeling,torunbalci2018modular}:
\begin{equation}
G_{MTJ}(V)  = G_0 \left[1+ P_1(V) P_2(V) \cos \theta_{1,2}\right]
\label{eq:cond}
\end{equation}
where $V$ is the bias voltage across the MTJ, $G_0$ is the conductance measured when $\cos\theta_{1,2}=\pi/2$ and $P_i(V)$ are the voltage dependent polarizations of the two interfaces. By a physically motivated choice of interface polarizations, this model reproduces the bias-dependence of the resistance, as well as the asymmetric bias dependence of the spin-polarized current, $\vec{I}_S(V)$ \cite{datta2011voltage}.  A reasonable model for the polarization is \cite{torunbalci2018modular}:
\begin{equation}
P(V)  = \frac{1}{1+ P_0 \exp(-V/V_0)}
\label{eq:pol}
\end{equation}
where $P_0$ is a parameter that is determined by the low-bias magnetoresistance and $V_0$ is determined by the high-bias features of the MTJ. This model is motivated by the observation that at higher voltages the polarization of the injected currents becomes weaker considering parabolic FM bands in contacts \cite{datta2011voltage}.

For simplicity, we assume a symmetric junction where $P_1(V)=P_2(-V)=P(V)$ is satisfied and we drop
the subscript to denote only one polarization function defined by Eq.~\ref{eq:pol}. We also assume a weak dependence of polarization
with respect to voltage by choosing a large roll-off parameter ($V_0$ ) compared to the applied biases of interest ( $\approx\pm 0.5$ V) for the p-bit considered in Section~\ref{sec:pbit}. 
In actual experiments, the bias-dependence of the torque can be controlled by the roll-off parameter that can introduce asymmetries in larger biases depending on the bias asymmetry of spin-polarized currents similar
to what is observed in standard MTJs \cite{datta2011voltage}.

With bias-dependent polarizations and conductance of the MTJ defined,
we can define the magnetization dependent spin-polarized current in the channel as \cite{datta2011voltage,camsari2014physics}:
\begin{equation}
\vec{I}_S (V) = G_0 V [P(V) \hat m_1 + P(-V) \hat m_2 ]
\label{eq:spincurrent}
\end{equation}
where the total spin-polarized current is the vectorial sum of two components proportional to the magnetization of each layer. In typical descriptions of spin-transfer-torque in MTJs, only one of these terms appear since the fixed layer is assumed inert. In the double-free layer system however both magnets are active  and they respond to spin-polarized currents that are polarized in the other magnets' direction, therefore we need to consider the total spin-polarized current.  A key point to note is that Eq.~\ref{eq:spincurrent} describes the total spin-polarized current in the channel. For one free layer, $+\vec{I}_s$ is incident to the ferromagnetic interface and for the other layer $-\vec{I}_s$ is incident to the interface. Therefore, what is supplied to LLG equations differ by a minus sign (FIG.~\ref{fi:model}). This can also be intuitively understood by considering one free layer as the instantaneously fixed reference layer of the other one: A +V ($-$V) bias that would make the layers parallel (antiparallel) would switch sign if we imagine the other magnet as the reference layer. 

There is also another term that is along a direction that is orthogonal to both $\hat m_1$ and $\hat m_2$, the so-called field-like torque, but it is typically small compared
to the main terms and we ignore it in this paper \cite{kubota2008quantitative,wang2011time,boyn2016twist}. The form of Eq.~\ref{eq:spincurrent} can be justified by microscopic quantum transport models based on the Non Equilibrium Green's Function (NEGF) formalism that is able to reproduce the bias-dependence of torque and resistance values in experiments \cite{datta2011voltage}. Even though there are two terms in Eq.~\ref{eq:spincurrent}, the individual magnetization dynamics of each free layer only picks up a torque from the transverse component of the other free layer \cite{PhysRevB.66.014407} since the form of Eq.~\ref{eq:llg} ensures that $\hat m_i  \times \vec{I}_S \times  \hat m_i$ cancels out components of $\vec I_S$ along $\hat m_i$.  We put together all the ingredients discussed so far, the dipolar tensors based on Eq.~\ref{eq:int}, the finite temperature LLG dynamics based on Eq.~\ref{eq:llg} and the transport equations of MTJ described by Eq.~\ref{eq:cond}, \ref{eq:pol}, \ref{eq:spincurrent} in a modular circuit environment \cite{torunbalci2018modular} that is simulated in HSPICE (FIG.~\ref{fi:model}). 

FIG.~\ref{fi:full_bias} shows representative voltage dependent characteristics of a double free layer MTJ.  A striking finding is that the bias dependence of the resistance is approximately independent of the applied voltage, even in the presence of the full effect of spin-transfer-torque between the layers. The bias dependence also shows symmetry with respect to voltage, a result we intuitively expect since the device is completely symmetric with two identical free layers. FIG.~\ref{fi:full_bias}a illustrates the dynamics of the free layers at different time instances. We observe that the free layers fluctuate close to the $x-y$ plane but also occasionally picking up a $z$-component. These fluctuations are reminiscent of the fast precessional fluctuations of easy-plane magnets that have been examined in Refs.~\cite{kaiser2019subnanosecond,hassan2019low,safranski2020demonstration} but all with stable reference layers unlike the case considered here. 

FIG.~\ref{fi:full_bias}b shows the R vs V characteristics of a 10 nm double-free layer MTJ where the resistance keeps fluctuating approximately uniformly between $R_P$ and $R_{AP}$ values at all bias voltages. This bias-independence of the fluctuations is a significant advantage of the double-free layer MTJ since it can provide a fluctuating resistance over a wide range of values without getting pinned, unlike MTJs with fixed layers \cite{borders2019integer,safranski2020demonstration} where the random fluctuation point needs to be identified carefully in an eventual device implementation. We also note from FIG.~\ref{fi:full_bias}c that above 60 to 100 nm, the average of angle (and the resistance of the MTJ), even though random, is largely stuck around the $R_{AP}$ value where the increasing dipolar energy with magnetic volume overcomes the thermal noise. For this reason, strategies to use scaled dimensions or low magnetic moment ($M_s$) materials will be useful. 

\begin{figure}[t] 
\centerline{\includegraphics[width=0.999\linewidth]{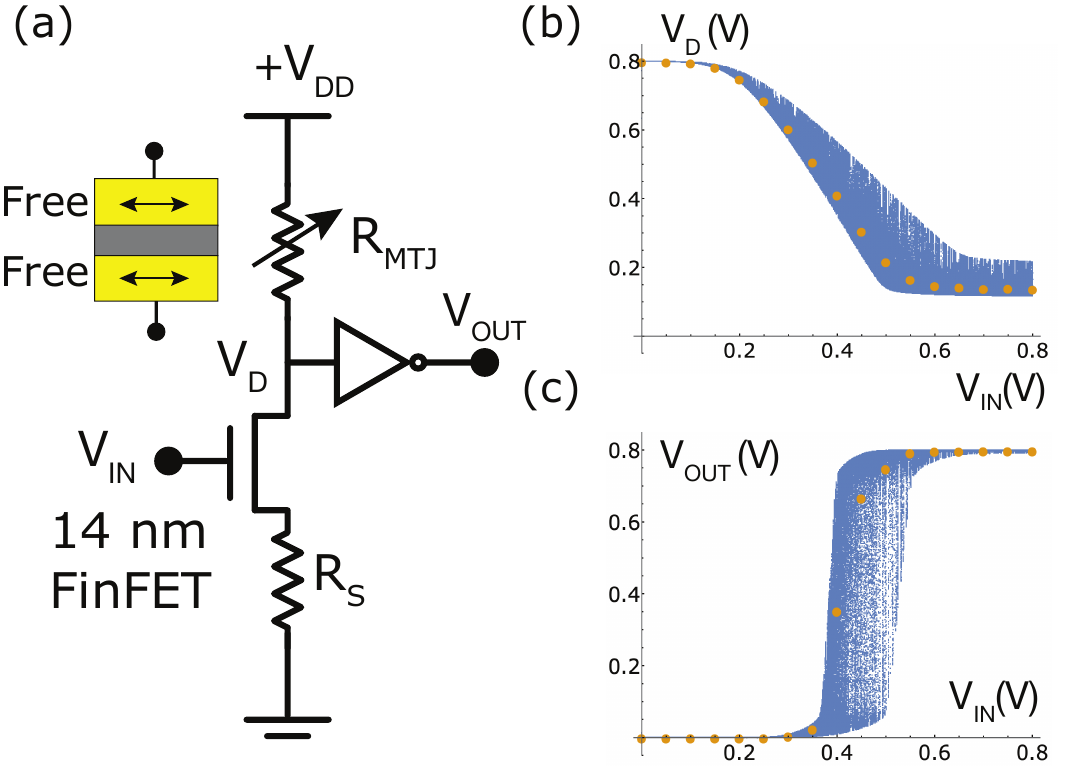}}
\caption{\textbf{Binary stochastic neuron with double-free layer MTJ:} (a)  The double-free layer MTJ ($2R=20$ nm with all the same parameters that are used in previous figures) in a 1T/1MTJ circuit, where a $R_S=5 \ k\Omega$ resistance is used to shift the overall characteristics. (b) The drain voltage is measured while the input is swept from 0 V to 0.8 V over 0.5 $\mu$s. The circled points are averages over 250 ns at each bias point. (c) The output of the inverter that shows binary stochastic neuron characteristics with the same measurement times reported in (b). }
\label{fi:pbit}
\end{figure}

\section{Tunable randomness with double free layer MTJ}
\label{sec:pbit}

In this section, we show how the double-free layer MTJ can be used to deliver  a hardware binary stochastic neuron (BSN) functionality in a 1T/1MTJ circuit plus an inverter circuit (FIG.~\ref{fi:pbit}) based on Ref.~\cite{camsari2017implementing} but we attach a source resistance $R_S$ to be able to shift the overall characteristics to the left or to the right similar to what was carried out in Ref.~\cite{borders2019integer}. We use a  High-Performance (HP) 14nm FinFET model from the Predictive Technology Model (PTM) \cite{predictive_tech} 
to model the NMOS and combine the NMOS model to our circuit model simulated in HSPICE. We choose the 20 nm MTJ (with the same parameters that were used in this paper) to illustrate the circuit operation (FIG.~\ref{fi:pbit}) since its average resistance (1/$G_0$) approximately matches the transistor resistance plus the source resistance ($R_S$) when the input to the NMOS is 0.4 V. Our purpose is not to provide a comprehensive circuit design of the BSN but simply to illustrate how the proposed MTJ that includes effects of thermal noise, dipolar coupling and bias-dependent spin and charge currents can be used to build a viable hardware BSN.

FIG.~\ref{fi:pbit} shows the output of the inverter while the input gate voltage of the NMOS is swept from 0 to $V_{DD}$ whose average shows the familiar sigmoidal behavior of the binary stochastic neuron. One potential challenge in the design of double-free layer MTJs will undoubtedly be designing the average P/AP ratio of fluctuations. Indeed, as can be observed from FIG.~\ref{fi:full_bias}c, the average resistance of the MTJ at different diameters is not in the middle of $R_P$ and $R_{AP}$ values but closer to $R_{AP}$ due to the dipolar coupling of the in-plane free layers. While at very high diameters $>$ 60 nm, this ratio may mostly be skewed toward $R_{AP}$, the skew at 20 nm can be mitigated by a source resistance ($R_S$) to center the sigmoid of the hardware BSN as shown in FIG.~\ref{fi:pbit}. We observe that changing this resistance value causes shift of the overall characteristics (with higher values causing a rightward shift). 

Unlike previous stochastic MTJs where the bias point where fluctuations between $R_{AP}$ and $R_{P}$ can significantly vary between different devices, the weak bias dependence of the proposed MTJ can be globally centered by a fixed source resistance.  This difference between having to align each p-bit precisely at their midpoint by a different bias current and obtaining an approximately uniform randomness at all relevant bias voltages constitutes an important advantage of the proposed design that can be exploited at the system level. Secondary variations arising from differences in MTJ resistances and transistor process variations can further be dealt with at the ``synaptic level'' where weighted inputs of p-bits can be modified by constant biases to counter these variations.  Further, probabilistic computations are generally robust to small variations \cite{pervaiz2017hardware,drobitch2019reliability} and strategies that may counter such variations by adjusting the interconnection weights of probabilistic devices (as was done in Ref.~\cite{borders2019integer}) can be useful.

Moreover, ultra-scaled MTJs beyond 20 nm can become truly random as predicted by Eq.~\ref{eq:dipolarA} provided that we can match the resistance of the NMOS to that of the MTJ, either by transistor design or by an additional source resistance as we have shown in FIG.~\ref{fi:pbit}. Device level simulations have shown that BSN characteristics similar to those shown in FIG.~\ref{fi:pbit} can be used as building blocks to design probabilistic circuits to solve optimization \cite{sutton_intrinsic_2017,camsari2017implementing,hassan2019voltage} and sampling problems to train neural networks \cite{kaiser2020probabilistic}. Therefore, the proposed double-free layer MTJ can function in similar ways to be useful for these applications. 

It is important to note that even though we have presented the double-free layer MTJ in terms of circular disk magnets that can fluctuate in nanosecond timescales, recent experimental \cite{safranski2020demonstration,hayakawa2021} and theoretical work \cite{kaiser2019subnanosecond,hassan2019low,kanai2021} have now firmly established that elliptical in-plane easy axis magnets can also fluctuate
in similar timescales. While the exact details of our model might be different, the double-free layer concept as a bias-dependent building block to build probabilistic bits applies to such structures with qualitatively similar results. 

\section{Conclusion}
We have proposed and analyzed a new magnetic tunnel junction with two free layers to generate random fluctuations using a comprehensive model where the dipolar interaction, thermal noise and bias-dependent spin and charge currents have been considered. Our findings reveal an approximately bias-independent magnetization fluctuations that can produce random resistance values at a wide range of bias values which can simplify circuit design with such MTJs. Another key advantage of the proposed magnetic stack is in its simplicity: The two free layers can be completely symmetric and their in-plane magnetization can easily be achieved by modifying the existing STT-MRAM technology by increasing the thickness of both free and fixed layers. Repurposing STT-MRAM technology with double-free layer MTJs  can lead to massively parallel tunable random number generators that can find useful applications in probabilistic computing.  

\section*{Acknowledgment}

Use was made of computational facilities purchased with funds from the National Science Foundation (CNS-1725797) and administered by the Center for Scientific Computing (CSC). The CSC is supported by the California NanoSystems Institute and the Materials Research Science and Engineering Center (MRSEC; NSF DMR 1720256) at UC Santa Barbara. KYC and MMT thank Daryl Lee from the Department of Electrical and Computer Engineering at UC Santa Barbara for providing computational resources during the course of this project. KYC gratefully acknowledges fruitful discussions with Jan Kaiser, Orchi Hassan and Supriyo Datta. The work is partly supported by JST-CREST JPMJCR19K3.

\end{document}